\documentclass[aps,twocolumn,prd,showpacs,nofootinbib]{revtex4}
\usepackage{amsmath}
\usepackage{graphicx}
\usepackage{dcolumn}
\usepackage{bm}
\usepackage{amssymb}
\usepackage{latexsym}

\def\be{\begin{equation}}
\def\ee{\end{equation}}
\def\ba{\begin{eqnarray}}
\def\ea{\end{eqnarray}}

\begin{document}

\title{Possible Explanation to Low CMB Quadrupole}

\author{Yun-Song Piao$^{a,b}$}
\affiliation{${}^a$Institute of High Energy Physics, Chinese
Academy of Sciences, P.O. Box 918-4, Beijing 100039, P. R. China}
\affiliation{${}^b$Interdisciplinary Center of Theoretical
Studies, Chinese Academy of Sciences, P.O. Box 2735, Beijing
100080, China}

\begin{abstract}

The universe might experience many cycles with different vacua.
The slow-roll inflation may be preceded by kinetic-dominated
contraction occurring in ``adjacent" vacua during some cycles. In
this report we briefly show this phenomenon may lead to a cutoff
of primordial power spectrum, which is mildly preferred by WMAP
data. Thus in some sense the CMB at large angular scale might
encode the information of other vacua.

\end{abstract}

\pacs{98.80.Cq} \maketitle

The interesting result of recent WMAP data,
which confirms earlier COBE observations \cite{CLB, Fang}, is a
lower amount of power on the largest scales when compared to that
predicted by the standard $\Lambda$CDM models \cite{Bennett,
Spergel}. This may be contributed to cosmic variance with bad
luck, where we might simply live in a region of universe with the
CMB quadrupole happening to be small. However this lower power
might also imply a cutoff of primordial power spectrum on the
largest scale \cite{Lewis}, which is related to the physics before
the onset of inflation \cite{Linde}. There are also many other
attempts \cite{Lsmall, PFZ, PTZ} to explain WMAP data.

Recently, a large number of vacuum states in string theory has
received many attentions \cite{Douglas, KKLT}, see also Ref.
\cite{BP, FLW}. The space of all such vacua has been dubbed
landscape \cite{Susskind}. In some sense, the low energy
properties of string theory can be approximated by field theory.
Thus the landscape can also be described as the space of a set of
fields with a complicated and rugged potential, where the local
minima of potential are called the vacua. When this local minimum
is an absolute minimum, the vacuum is stable, and otherwise it is
metastable. In string landscape with exponentially large number of
vacua we can only live in one vacuum compatible to us, which might
make us able to anthropically solve the problem of cosmological
constant that has been troubling us for a long time \cite{W, MSW}.
Further it has been shown in Ref. \cite{Piao} that for a landscape
with a large number of AdS minima the universe may experience many
cycles \cite{KOS} with different minima, when the number of cycles
is large or approaches infinity, whichever minimum initially the
universe is in, it can run over almost all vacua of the landscape.
Thus in some sense the physics of adjacent vacua settles the
initial conditions and affects the evolution of universe with the
vacua observed. Further this might leave an observable imprint in
CMB under certain conditions.
We will briefly illustrate this possibility in this report.



For a simplified example given in terms of an
order parameter $\varphi$, the effective description of a
landscape with many AdS minima can be taken as follows \be {\cal
V}(\varphi) = \Lambda_*\left(1-\cos({m\over
\sqrt{\Lambda_*}}\varphi)\right)-\Lambda,\label{v}\ee where
$\Lambda$ is a small positive constant which makes the minima of
periodic potential negative, and $m$ is the mass around the minima
of potential. The universe with negative potential will eventually
collapse \cite{FFKL}. But Big Crunch singularity might be not a
possible feature of quantum gravity. There should be some
mechanisms from high energy/dimension theories responsible for a
nonsingular bounce. We suppose, following this line, that in high
energy regime the Friedmann equation can be modified as $ 3 h^2
\simeq \rho_{\varphi} -\rho_{\varphi}^2 / \sigma$ \cite{SS}, where
$8\pi /m_p^2 =1$ has been set, and $\sigma$ is the bounce scale.

Following Ref.\cite{Piao}, the universe, controlled by a rugged
potential with many AdS minima and a bounce mechanism in high
energy regime, will show itself many contraction/expansion cycles.
The functions of the field and scale factor with respect to time
are plotted in Fig. 1. When the bounce scale $\sigma$ is larger
than the height of potential hill, the field will be driven from a
minimum to another, and during each cycle of oscillating universe
the field will generally lie in different minima. The kinetic
energy of the field during contraction will rise rapidly and be
much larger than its potential energy, which makes the field able
to get over the potential barrier easily and quickly. The maximum
value to which the field is driven during each
contraction/expansion cycle can be simply estimated as \cite{Piao,
PZ} \be \varphi_m \simeq 1 + \sqrt{2\over 3}\ln{({\sigma\over m^2
})}, \label{phibou2}\ee which is only relevant with the mass $m$
around its minima and the bounce scale $\sigma$. For the potential
of (\ref{v}), in the apex of hill, $\varphi_a= {\pi
\sqrt{\Lambda_*}\over m}$. Thus to make the field $\varphi$ can
stride over a potential hill during a cycle, $\varphi_m \gtrsim
{\pi \sqrt{\Lambda_*}\over m}$ has to be satisfied.


\begin{figure}[t]
\begin{center}
\includegraphics[width=8.8cm]{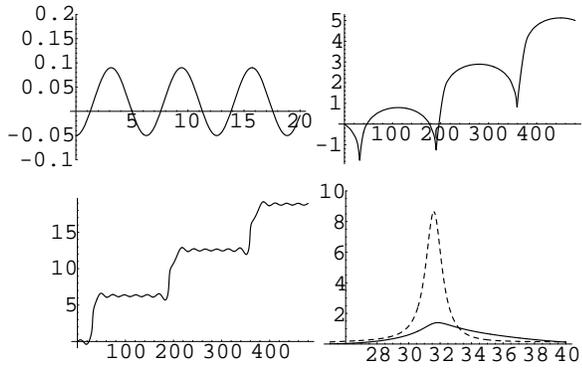}
\caption{The upper left panel is the figure of potential (\ref{v})
for illustration. The upper right panel is the figure of $\ln{a}$
with respect to time, where $\Lambda_* =0.07$, ${m\over
\sqrt{\Lambda_*}}=1$ and $\Lambda=0.0005$ are taken. The lower
left panel is the figure of the value of field with respect to
time, while lower right panel is the figure of its potential
energy (solid line) and kinetic energy (dashing line) in time
interval (25, 40). The field generally oscillates in different
minima during different cycles of universe.  }
\end{center}
\end{figure}

We assume that after getting across the potential barrier and
before rolling down to its new minimum, the field can enter into a
phase dominated by a flat part of potential, which drives a period
inflation of universe, see Fig. 2 for an illustration, where $m_1$
and $m_2$ are the mass scale of two adjacent minima respectively.
The potential at the right side of the apex $\varphi_a$ of
potential hill can be written as $2(\Lambda_* -{1\over
2}m^2_2\varphi^2)$, which may be regarded as an expansion of Eq.
(\ref{v})-like potential around $\varphi_a$. The e-folds number
during the inflation for above potential is \be {\cal N}= \int h
dt \simeq {\Lambda_*\over m_2^2}\ln({\varphi_e\over \varphi_i}) ,
\label{n}\ee where $\varphi_e$ is the value of field in which the
inflation ends, which can be approximately given by the slow-roll
parameter $\epsilon \simeq {m_2^4 \varphi^2_e\over \Lambda_*^2}
\simeq 1$, thus $\varphi_e \simeq {\Lambda_* \over m_2^2}$, and
$\varphi_i$ is the value of field in which the inflation begins,
which is determined by the physical parameters of last minimum,
\be \varphi_i= \triangle\varphi\simeq \varphi_m - {\pi
\sqrt{\Lambda_*}\over m_1} . \label{tri} \ee From Eq. (\ref{tri}),
we can see that in principle the physics of adjacent minima can
determine the possibility and e-folds number of inflation in
succedent minimum.
For $\varphi_i= \triangle\varphi < \varphi_e$, the inflation will
occur, and after the end of inflation, the parameteric resonance
\cite{KLS} of inflaton will lead to the production of a large
number of particles. The decay of these particles will reheat the
universe to required temperature, and then the standard
cosmological evolution begins. The universe will recollapse and
enter into next cycle until the energy density of matter is equal
to that of AdS minimum.

\begin{figure}[t]
\begin{center}
\includegraphics[width=7cm]{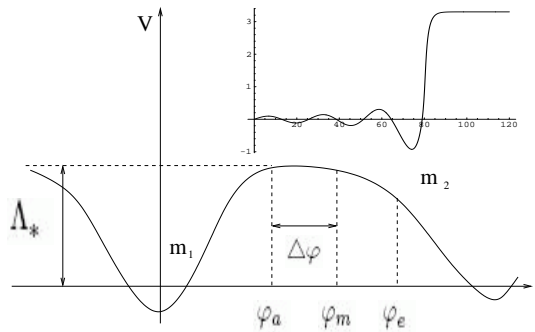}
\caption{ The illustration of model, where $m_1$ and $m_2$ are the
mass scale of two adjacent vacua respectively, and $\varphi_i$ is
the value of field at the onset of inflation and $\varphi_e$ is
the value of field in which the inflation ends. The inset is the
figure of the value of field with respect to time during a
contraction/expansion cycle, where the effective potential is
taken as a combination of potential of (\ref{v}) and a flat
potential, which are matched at $ \varphi_a= {\pi
\sqrt{\Lambda_*}\over m}$, and $\Lambda_* =0.07$, ${m\over
\sqrt{\Lambda_*}}=1$ and $\Lambda=0.001$. Initially the field
oscillates in its minima with mass scale $m_1$, then roll up along
its potential, and after the bounce of universe, the field enters
into the slow-roll regime rapidly, in which the mass scale is
$m_2$. }
\end{center}
\end{figure}

We then calculate the primordial spectrum of the above model
discussed. There are generally two regimes for the generation of
primordial spectrum in this model, namely kinetic-dominated phase
and succedent inflationary phase. For simplify, we neglect the
details of the bounce and pay much attention to an instantaneous
transition between both phases. Following \cite{PFZ}, the scale
factors of two different phases can be given by \be a\simeq
\sqrt{1-2{\cal H}_0\eta}~, ~~~~~\eta\leq 0 \label{leq} \ee \be
a\simeq {1\over 1-{\cal H}_0\eta}~, ~~~~~\eta\geq 0 \label{geq}
\ee respectively, where $\eta$ is the conformal time, $\eta=0$ and
$a=1$ have been set for the matching at the moment of transition,
and ${\cal H}_0$ is the value of ${\cal H} ={a^\prime \over a}$ at
$\varphi=\varphi_i$, in which the universe just enters into
inflationary phase.

The variable \cite{M} \be v\equiv a\left(\delta \varphi
+{\varphi^\prime\over {\cal H}}\Phi\right)\equiv z\zeta ,\ee is
defined for the calculations of perturbation spectrum, where
$\Phi$ is the Bardeen potential \cite{B} and $\zeta$ is the
curvature perturbation on uniform comoving hypersurface, and
$\delta \varphi$ is the perturbations of the scalar field during
both phases, and $z\equiv {a\varphi^\prime\over {\cal H}}$, see
\cite{KS, MFB} for a thorough introduction to gauge invariant
perturbations. In the momentum space, the equation of motion of
$v_k$ is \be v_k^{\prime\prime}+\left(k^2 -{z^{\prime\prime}\over
z}\right)v_k =0  .\ee For the kinetic-dominated contracting phase,
\be {z^{\prime\prime}\over z}\simeq {a^{\prime\prime}\over a}
\simeq {-{\cal H}_0^2\over (1-2{\cal H}_0 \eta)^2} . \ee When
$k^2\gg {z^{\prime\prime}\over z}$, the fluctuations are deep in
the contracting phase and can be taken as an adiabatic vacuum,
which corresponds to \be v_k\sim {1\over \sqrt{2k}} e^{-ik\eta}
,\ee thus \be v_k(\eta)=\sqrt{\pi(1-2{\cal H}_0\eta)\over 8 {\cal
H}_0}{\cal H}_0^{(2)}\left(-k\eta+{k\over 2{\cal H}_0}\right)
 ,\ee where ${\cal H}_0^{(2)}$ is the second kind of Hankel function
with $0$ order. For the nearly de Sitter phase, \be
{z^{\prime\prime}\over z}\simeq {a^{\prime\prime}\over a} \simeq
{2{\cal H}_0^2 \over (1-{\cal H}_0\eta)^2} ,\ee thus \ba & &
v_k(\eta)=\sqrt{-k\eta +{k\over {\cal H}_0}}\nonumber \\ &
&\left(C_1 {\cal H}_{3\over 2}^{(1)}(-k\eta +{k\over {\cal
H}_0})+C_2 {\cal H}_{3\over 2}^{(2)}(-k\eta +{k\over {\cal
H}_0})\right) \label{vki} ,\ea where ${\cal H}_{3\over 2}^{(1)}$
and ${\cal H}_{3\over 2}^{(2)}$ are the first and second kind of
Hankel function with ${3\over 2}$ order respectively, $C_1$ and
$C_2$ are the functions dependent of $k$, and are determined by
the matching conditions between both phases, which are related to
the physics around the bounce
and specifically depend on which of $\zeta$ and $\Phi$ passes
regularly through the bounce \cite{CDC}. The continuities of $v$
and $v^\prime$ \cite{vvp} at the transition give
 \ba C_1 =&
&\sqrt{\pi\over 32{\cal H}_0}e^{-ik\over {\cal H}_0} ((1-{2{\cal
H}_0^2\over k^2}-{2{\cal H}_0\over
k}i){\cal H}_0^{(2)}\left({k\over 2{\cal H}_0}\right)\nonumber \\
& & +({{\cal H}_0\over k} +i){\cal H}_1^{(2)}\left({k\over 2{\cal
H}_0}\right)) ,\label{c1}\ea \ba C_2 &=& \sqrt{\pi\over 32{\cal
H}_0}e^{ik\over {\cal H}_0} ((1-{2{\cal H}_0^2\over
k^2}+{2{\cal H}_0\over k}i){\cal H}_0^{(2)}\left({k\over 2{\cal H}_0}\right)\nonumber\\
& &+({{\cal H}_0\over k} -i){\cal H}_1^{(2)}\left({k\over 2{\cal
H}_0}\right)) ,\label{c2}\ea where ${\cal H}_0^{(2)}$ and ${\cal
H}_1^{(2)}$ are the second kind of Hankel function with $0$ and
$1$ order respectively.

The perturbation spectrum is \be {\cal P}_s = {k^3\over
2\pi^2}|{v\over a}|^2 \label{p} \ee for $\eta \rightarrow 1/{\cal
H}_0$. Substituting (\ref{vki}), (\ref{c1}) and (\ref{c2}) into
(\ref{p}), we obtain \be {\cal P}_s= {{\cal H}_0^2\over
2\pi^2}k|C_1 -C_2|^2. \label{ps}\ee For $k\ll {\cal H}_0$, the
Hankel function can be expanded in term of large variable, thus we
obtain approximately $P_s \sim k^3$ on large scale, which is the
usual result of PBB scenario \cite{GV}. For $k \gg {\cal H}_0$,
the expansion of small variable of the Hankel function gives $P_s
\sim k^0$ {\it i.e.} nearly scale-invariant spectrum on small
scale.
The result of numerical calculation is plotted in inset of Fig. 3,
where the cutoff of spectrum can be seen clearly, which is
consistent with above semianalytical ones. The large $k$ modes are
generally inside the horizon during the kinetic-dominated phase
and are not quite sensitive to the background at this stage. Thus
when they cross the horizon during inflation after the transition,
the nearly scale-invariant spectrum can be generated by the
evolution of the background during inflationary phase.

\begin{figure}[t]
\begin{center}
\includegraphics[width=8.8cm]{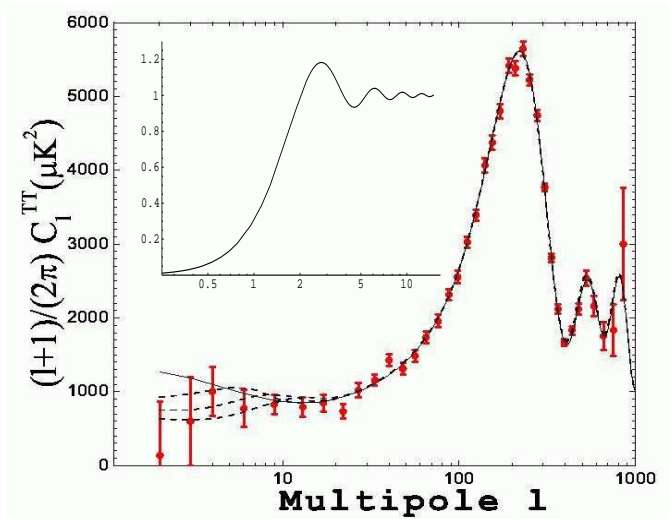}
\caption{  The CMB anisotropy for the scale-invariant spectrum and
the spectrum with a cutoff. From left top to bottom, the lines
stand for scale-invariant spectrum, spectrum with a cutoff with
${\cal H}_0=2.1,3.1$ and $4.1\times 10^{-4}$ Mpc$^{-1}$. The other
parameters are fixed at $h=0.73$, $\Omega_b h^2=0.023$,
$\Omega_{cdm} h^2= 0.117 $ and $\tau=0.2$. The inset is the power
spectrum ${\cal P}_s$as a function of ${k\over {\cal H}_0}$. The
x-axe is ${k\over {\cal H}_0}$, and the y-axe is ${\cal P}_s
/\left({{\cal H}_0\over 2\pi}\right)^2$.}
\end{center}

\end{figure}

We fit the resulting primordial spectrum to the current WMAP data.
In Fig. 3, we show the CMB TT multipoles for the scale-invariant
spectrum and the cutoff spectrum with various ${\cal H}_0$.
Regarded as a cutoff scale in the spectrum, ${\cal H}_0$ can be
chosen as ${\cal H}_0\lesssim 5.0\times 10^{-4}$ Mpc$^{-1}$ in our
fit. From Fig. 3, we see that the lower CMB TT quadrupole is
related to the value of ${\cal H}_0$, which can be determined by
Eq. (\ref{tri}). Thus in some sense lower CMB quadrupole encodes
the information of adjacent vacua. In addition we get a minimum
$\chi^2=1428.2$ at $h=0.73$, $\Omega_b h^2=0.024$, $\Omega_{cdm}
h^2= 0.116 $, $\tau=0.2$ and ${\cal H}_0=2.0\times 10^{-4}$
Mpc$^{-1}$. We also run a similar code for the scale-invariant
spectrum for comparison and get a minimum $\chi^2=1429.7$ at
$h=0.73$, $\Omega_b h^2=0.024$, $\Omega_{cdm} h^2= 0.116 $ and
$\tau=0.2$. This means the primordial spectrum in our example is
favored at $>1.22\sigma$ than the scale-invariant spectrum.

We notice that the errors in power spectrum estimates, especially
at low l, are highly non-Gaussian, thus the model depends in a
nonlinear way on the parameters, which to some extent makes our
simply numerical analysis incorrect. To assign a statistical
significance correctly, either Monte Carlo simulations or a full
Bayesian analysis is necessary. However, this brief report is not
primarily about these statistical issues, thus a
Numerical-Recipes-level analysis may be enough.


Though our model does not solve the problem of the low CMB
quadrupole completely, it provides a mechanism leading to the
cutoff of primordial power spectrum, which in some sense is mildly
preferred by WMAP data. 
The suppression of CMB quadrupole in our model
is significantly dependent of parameters of adjacent minima.
However, since the number of vacua in the landscape is
exponentially large, there may always exist some adjacent vacua
with such characters, thus the probability that an observer finds
a suppression with intension observed will be not too small, which
in some sense relaxes the requirement for fine-tuning.



Facing diverse vacua \cite{Susskind} to which string theory brings
us, what people might very long for is ``seeing" adjacent or other
parts of landscape. Thus trying to read some information of other
parts from observations will be an excited thing. Though the
example in this brief report may be idealistic and speculative,
it might identified some of the basic ingredient of required
answer. We leave the realistic implementations \cite{Piao2}
and other interesting applications to future works.

{\bf Acknowledgments} The author thank Bo Feng for helpful
discussions and kindly offer of CMB anisotropy figure (Fig.3).
This work is supported in part by K.C. Wang Posdoc Foundation,
also in part by NNSFC under Grant No: 10405029 , 90403032 and by
National Basic Research Program of China under Grant No:
2003CB716300 .

\end{document}